\documentclass[acmlarge]{acmart}

\AtBeginDocument{
  \providecommand\BibTeX{{
    \normalfont B\kern-0.5em{\scshape i\kern-0.25em b}\kern-0.8em\TeX}}}

\usepackage[utf8]{inputenc}
\usepackage{svg}
\usepackage{siunitx}
\usepackage{xspace}
\usepackage{url}

\usepackage{breakurl}
\usepackage{array}

\makeatletter

\newcolumntype{"}{@{\hskip\tabcolsep\vrule width 1.4pt\hskip\tabcolsep}}
\makeatother

\usepackage{longtable}

\newcommand{\RQBOX}[1]{
  \noindent\fbox{\parbox{0.97\linewidth}{\bfseries
      #1
    }}}

\newcommand{\figref}[1]{\figurename~\ref{#1}}

\setcopyright{acmcopyright}
\copyrightyear{2020}
\acmYear{2020}
\acmDOI{10.1145/1122445.1122456}

\acmConference[ICMLC 2021]{International Conference on Machine Learning and Computing 2021}{26 Feb-1 Mar, 2021}{Shenzhen, China}
\pdfmapfile{+txfonts.map}

\title{Content-Based Textual File Type Detection at Scale}

\author{Francesca Del Bonifro}
\email{francesca.delbonifro@unibo.it}
\affiliation{
  \institution{University of Bologna}
  \city{Bologna}
  \country{Italy}
}

\author{Maurizio Gabbrielli}
\email{maurizio.gabbrielli@unibo.it}
\affiliation{
  \institution{University of Bologna}
  \city{Bologna}
  \country{Italy}
}

\author{Stefano Zacchiroli}
\email{zack@irif.fr}
\orcid{0000-0002-4576-136X}
\affiliation{
  \institution{Université de Paris and Inria}
  \city{Paris}
  \country{France}
}

\begin{abstract}
  Programming language detection is a common need in the analysis of large
  source code bases. It is supported by a number of existing tools that rely on
  several features, and most notably file extensions, to determine file types.
  We consider the problem of accurately detecting the type of files commonly
  found in software code bases, based solely on textual file content. Doing so
  is helpful to classify source code that lack file extensions (e.g., code
  snippets posted on the Web or executable scripts), to avoid misclassifying
  source code that has been recorded with wrong or uncommon file extensions,
  and also shed some light on the intrinsic recognizability of source code
  files.
  We propose a simple model that (a) use a language-agnostic word tokenizer for
  textual files, (b) group tokens in 1-/2-grams, (c) build feature vectors
  based on N-gram frequencies, and (d) use a simple fully connected neural
  network as classifier. As training set we use textual files extracted from
  GitHub repositories with at least 1000 stars, using existing file extensions
  as ground truth. 
  Despite its simplicity the proposed model  reaches $\approx$\,85\%
  in our experiments for a relatively high number of recognized classes (more than 130 file types).
\end{abstract}

\begin{document}
\maketitle

\section{Introduction}
\label{sec:intro}

\emph{``Which programming language is it written in?''} is among the first
questions a developer would ask about any piece of software. In software
engineering research, ``programming language'' is a common variable to
correlate against---researchers have measured language
productivity~\cite{maxwell1996productivity, maxwell2000productivity,
  rodriguez2012productivity}, trends~\cite{rabai2011trends,
  DBLP:conf/ecoop/2014ple}, usability~\cite{mciver2000novices}, and code
quality~\cite{kochhar2016codequality, ray2017codequality,
  vitek2019codequality}, to name just a few. Developers would know by heart the language their code is written in, and will
also easily identify the main programming language used in any given code base.
But when it comes to \emph{large} code bases, which often mix and match
multiple programming languages; or when it comes to large-scale
analyses~\cite{dyer2013boa, mockus2019woc}, encompassing entire collaborative
development forges if not the full body of publicly available source
code~\cite{swh-cacm-2018}, manual detection is not an option. In those cases
researchers tend to rely on either off-the-shelf language detection tools
(e.g., GitHub's Linguist~\cite{linguist}), or on metadata exposed by code
hosting platforms~\cite{librariesio2018}, which were in turn extracted using
the same tools.

File \emph{extensions} are highly predictive of programming languages and, more
generally, file types. As such detection tools tend to heavily rely on them to
determine file types~\cite{linguist-accuracy}. While very effective in the
general case, doing so is problematic when file extensions are either not
available (e.g., code snippets posted on the Web or embedded in document,
executable scripts, etc.) or wrong (either on purpose or by mistake).

In this paper we study the problem of detecting the file type of textual files
commonly found in software source code. Specifically, we will answer the
following research question:

\RQBOX{
  RQ: is it possible to accurately detect the extension of textual files
  commonly found in software version control systems repositories, based solely
  on file contents?
}

We will interpret ``solely'' in a strict sense, depriving ourselves of the use
of any \emph{a priori} heuristic on the content of the files that will be
encountered---i.e., our detection approach will \emph{not} know a priori the
keywords in the grammar of any given programming language, nor how shebang
(\texttt{\#!/usr/bin/perl}) or editor mode lines (\texttt{\%\%~mode:~latex})
are usually written in files. This is important for several reasons. First, programming languages (which are
the most significant part of textual file types encountered in code bases)
evolve and diverge syntactically significantly over time
\cite{wexelblat1981hopl-i, bergin1996hopl-ii, ryder2007hopl-iii,
  sammet1972langhist,DBLP:series/utcs/GabbrielliM10}. As it
interesting~\cite{shustek2006swpreservation} and is now becoming increasingly
more possible~\cite{swh-ipres-2017} to analyze historically relevant software
source code dating back several decades, heuristics built today will fail on
old code and heuristics spanning several generations of file formats will be
fragile and hard to maintain in the long run.  Second, extensionless file type
detection shed some light on the intrinsic recognizability of textual file
types, and programming languages in particular, which is an interesting topic.

As a consequence of the above we refuse to use as ground truth about file types
the notion of programming language---which, as we have seen, is actually not
well-defined in datasets spanning several decades, or that might contain
syntactically invalid files, files that mixes multiple programming languages,
etc; also, we did not want to rely on any pre-existing language detection tool
to not negatively compound accuracy.  Instead, we use file extensions
encountered in real code bases as file type classes and attempt to detect them
based on file content only. To stress this important point: file extensions
will only be used as labels in training sets and to measure accuracy, but not
as file features in the detection phase (as that would replicate the weakness
of current tools that we want to address).

We propose a simple model for the task at hand: (a) use a language-agnostic
tokenizer for textual files, (b) group tokens in n-grams of length 1 to 2, (c)
build feature vectors based on n-gram frequencies, and (d) use a simple fully
connected neural network as the actual classifier. As initial training set we
use $\approx$\,16\,M textual files extracted from popular GitHub repositories
(as determined by ``stars''~\cite{borges2018ghstars}), using associated file
extensions as labels; we further cleanup the dataset by filtering out binary
files and downsampling classes so that the number of samples per class are
analogous. In spite of its simplicity, the proposed classification approach reach an
aggregate precision of 91\% on a total of 133 file type classes, outperforming
state-of-the-art tools and approaches in precision, amount of considered
classes, or both. Thanks to its simplicity, the model is also fast to
(re)train and deploy, making the proposed approach more maintainable
in the long run.
The remaining of the paper is organized as follows. In Section~\ref{sec:dataset} we present the experimental dataset. The proposed
classification model is introduced in Section~\ref{sec:model} and
experimentally validated in Section~\ref{sec:results}. Threats to validity are
discussed in Section~\ref{sec:threats}. Comparisons with related work, both
qualitative and quantitative, are given in Section~\ref{sec:related}. We
discuss future work and wrap up the paper in Section~\ref{sec:conclusion}.
A complete replication package for this paper is available from Zenodo at
\url{https://zenodo.org/record/3813163} (DOI:
\href{http://dx.doi.org/10.5281/zenodo.3813163}{10.5281/zenodo.3813163}).

\section{Dataset}
\label{sec:dataset}

As experimental dataset we want files coming from public version control
system. We retrieved a snapshot of GitHub dated 2017-01-27 and made available
by the Software Heritage~\cite{swh-cacm-2018, swh-ipres-2017}
project,\footnote{\url{https://annex.softwareheritage.org/public/dataset/content-samples/2017-01-27-github-1000+stars/}}
which contains all source code files extracted from all commits of GitHub
projects ranked with 1000 or more ``stars''~\cite{borges2018ghstars}. It is a
significant dataset, weighting 141\,GB after compression and file-level
deduplication, containing $\approx$15\,M unique files. Each included file is identified by a SHA1 cryptographic checksum; a mapping
between (unique) files and the corresponding file names as encountered on
GitHub is provided. Due to deduplication it might happen that the same SHA1 is associated to
different filenames, associating it to different extensions.  The number of
different extensions in the mapping is $\approx$546\,K. However looking at
extension frequencies in the dataset it is immediately clear that many of them
are either typos or uncommon extension assignments. We set a threshold of
$10^{-4}$ for extension frequencies in order to filter out non statistically
significant extensions, thus obtaining 220 extensions, still remaining in
the $\approx$15\,M ballpark for number of files, confirming the validity of the
selection criteria.

A second filtering step consisted in excluding binary files, as our main
purpose for this paper is classifying textual files. As criterion for this we
have considered files as a sequence of raw bytes (as no explicit encoding
information is stored by Git), put a threshold of 20\% printable characters,
and excluded all files (and associated extensions) with more non printable
characters than that. After this filtering step we obtained 133 different
extensions and $\approx$13\,M files. A very small percentage of this set (less than 0.4\%) consists of files
associated to more than one extension. Rather than dealing with multi-label
classification, and given its low occurrence, we have excluded multi-extension
files from the dataset. We have obtained this way a multi-class, single-label
classification problem with a dataset of \num{12903304} files and 133 classes.

\begin{figure}
\begin{minipage}[]{10cm}
\centering
\includegraphics[width=9.0cm]{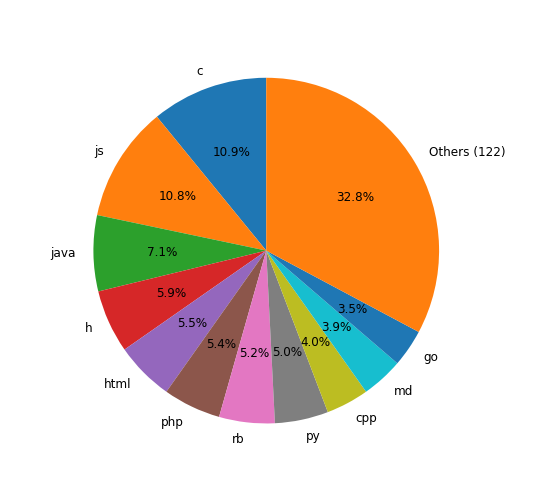}
\caption{Extension distribution in our corpus.}
\label{fig:extdist}
\end{minipage}
\ \hspace{0mm} \hspace{0mm} \
\begin{minipage}{4cm}
\centering
\includegraphics[width=4cm]{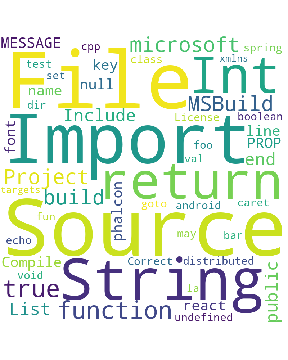}
\caption{Word cloud for token frequency distribution.}
 \label{fig:wordcloud}
\end{minipage}
\end{figure}

Most of the extensions in the polished dataset are commonly used as extensions
for programming or markup languages (the extensions with the highest
frequencies were \verb py , \verb rb , \verb html , \verb po , \verb php ,
\verb h , \verb java , \verb js , \verb c ), but the dataset also contains
extensions typically associated to textual files of other nature. Moreover, even though we ignored rare and weird extensions, the frequency range
remains wide, with some classes containing many examples while others just a
few, as shown in~\figref{fig:extdist}).  This represents a serious training
issue for most the supervised learning approaches, since models obtained with
unbalanced datasets tend to overfit and do not generalize
well~\cite{DBLP:books/lib/Bishop07}. In the following we describe how we solved
address this problem and built a more balanced and fair train set.

\subsection{Balancing the dataset}
\label{sec:balance}
The dataset as obtained thus far is not balanced, since the most frequent
extension (\verb c ) has a frequency of $10^{-1}$ while the least frequent one
(\verb tcl ) has a frequency of $10^{-4}$. Several techniques can be used for dataset balancing, and they are usually
divided in two major classes: oversampling and
undersampling~\cite{DBLP:reference/dmkdh/2010}. In the first case, new
instances for the minority classes are generated until they reach a population
similar to those of majority classes. The new instances are either copies of
the existing examples or synthesized by using statistical properties computed
on the minority classes.
In our case, since the unbalancing is very significant, we could be forced to
use the same examples (or the same information extracted from few examples) too
many times, increasing the risk of overfitting. We did not consider the option
of synthesizing artificial samples to avoid introducing biases, given
understanding how intrinsically recognizable are \emph{real} textual files
found in VCS is part of our goal. Rather, we have applied an undersampling technique consisting in (sub)sampling
the various classes randomly to obtain the same number of elements for each
class. Of course, with this approach we are limited by the number of instances
of the less populated class. As a result we ended up with a train set
containing \num{127300} total examples and in which each class has the same
number of instances ($\approx$950). In addition to the training set we also kept, as it customary, a test set and a
validation set, these sets preserve the original dataset classes distribution in order to represent a real situation.

In order to check model resistance to VCS evolution after a few years we have
also used a second, more recent, test set, consisting of new files which do not
appear in the first dataset, corresponding to the top 1000 GitHub repositories
as archived by Software Heritage on
2019-10-08\footnote{\url{https://annex.softwareheritage.org/public/dataset/content-samples/2019-10-08-github-top1k/}}.
The second dataset set was prepared 
with the same methodology used
for the first one, and contained 121 of the previously treated 133 extensions.

\section{The model}
\label{sec:model}

\begin{figure}
\includegraphics[scale = 0.65]{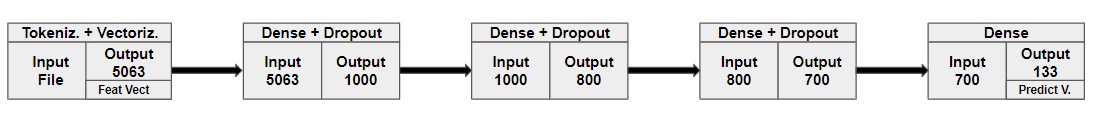}
  \caption{File type classification model.}
  \label{fig:model}
\end{figure}

The proposed classification model is developed in several steps. first the
content of the input file is divided into tokens which are then used to define
a reference vocabulary \emph{V}. Then, as common practice in many NLP
applications, we also construct of a vocabulary \emph{$V_2$} containing
2-grams: These are sub-sequences of 2 tokens extracted from the files, once
these are ``tokenized'', that is represented as sequences of tokens according
to the vocabulary \emph{V}.  At this stage we can extract the actual feature vectors
from the files by considering the frequency of words and of 2-grams in the file
text\footnote{Not all possible words and 2-grams are considered, of course, but
  only those appearing in the vocabularies \emph{V} and \emph{$V_2$},
  respectively.}. Finally the classifier which makes predictions on the basis
of the feature vectors is defined, by using a simple neural network whose
structure is depicted in Figure~\ref{fig:model}); these steps are
detailed in the remainder of this section.

\subsection{Tokenization}

Since our files are available as sequences of bytes, to treat them as texts we
first need to decode them using a suitable character encoding.  We used the
ASCII encoding and considered only the characters which are the most common in
source code. All non-ASCII characters are mapped to an unique, special value.
After this conversion we can consider our files as simple text files, i.e.,
sequences of characters, on top of which we can define a notion of token.
Differently from what is usually done in Natural Language Processing (NLP),
case sensitivity is relevant in our setting, hence preserve character
case-ness. Also, while in NLP punctuation symbols are discarded, they are
crucial in source code, so we consider them as tokens.\footnote{With the
  exception if \texttt{\_} which is considered an alphanumeric character, as it
  is often part of identifiers in source code.} Hence the following:
\begin{definition}
  Given a sequence of characters $S$, a \emph{token} (or equivalently a
  \emph{word}) in $S$ is defined as follows:
  \begin{itemize}
  \item Any character representing a punctuation symbol is a token\;
  \item Any sub-sequence of $S$ which is delimited by (characters representing)
    punctuation symbols and/or white spaces, and which does not contain
    punctuation symbols and/or white spaces is a token.
  \end{itemize}
\end{definition}

Thus, for example, the string `a=b' is interpreted as a sequence of the three tokens `a',
`=' and `b'. 
For model manageability we cannot consider all the possible words that occur in
any file. A common technique for addressing this is looking at the frequencies
of the tokens that occur in the trainset, and assemble a vocabulary consisting
of all the tokens whose frequency is higher than a given threshold. However,
due to heterogeneity of our dataset, this approach could cause the exclusion of
tokens that are quite common only for specific classes, thus causing poor
performance.\footnote{This was experimentally verified on our corpus} Hence we
computed the frequencies \emph{for each class} and included in the vocabulary
the tokens with frequency higher than $10^{-2}$ for each class.
To mitigate overfitting risks we have defined the vocabulary \emph{V} only by
using a part of the trainset (still a balanced set), which was then excluded
from the set used to train the network~\cite{singhi2006feature}. Many files in code bases include at their beginning and/or end explicit
information about the file content, in the form of shebang lines (e.g.,
\texttt{\#!/usr/bin/perl}) or editor mode lines ((\texttt{\%\%~mode:~latex})).
This information can be really helpful for the classification task, but it can
also compromise the performance the model, since this information could gain
too much relevance with respect to other features, inducing poor results on
files which lack it. For this reason, when collecting tokens to build the
vocabulary and during training we excluded a portion of tokens from both the beginning and end of
files.

The resulting
vocabulary \emph{V} contains 465 tokens, which are represented with their
own identity. The special token ``UNK'' represents any unknown,
out-of-vocabulary (OOV) token. Figure~\ref{fig:wordcloud} contains a word cloud
(where word size is proportional to the frequency of the token in the dataset)
representation of the tokens in the final vocabulary except for punctuation symbols.

\subsection{n-grams}
\label{sec:n-grams}

Information about the relative position of tokens in a text can be richer than information about isolated tokens. There exist
various algorithms and techniques that can capture different kind (e.g., short
or long) of relation among tokens. Some of them like, CNN and
LSTM~\cite{wenpeng2017comparative, yang2016hierarchical}, can capture
meaningful relations, automatically but they are also computational quite
expensive. Simpler approaches are based on n-grams, i.e., sub-sequences of $n$ tokens
extracted from a sequence of tokens defined according to a given vocabulary
\emph{V}. Taking into account n-grams, instead of individual tokens, it is
possible to identify \emph{co}-occurrences of tokens, extracting more
information about the actual text structure. By increasing the n-grams length
(the value $n$) it becomes possible to capture longer and more complex
relations among tokens, at the cost of increased computational costs and
increased overfitting risk---since longer n-grams tend to become tightly bound
to the text they are derived from. For this work we have used bigrams, i.e.,
$n=2$, which turned out to be a good choice in the performance/overfitting
spectrum. We have also experimented models with trigrams which have worse performance.

To define the bigram vocabulary we relied on the same approach used for
building the token vocabulary \emph{V}. We define \emph{$V_2$} as the set of
all bigrams whose frequency is higher than $10^{-3}$) in each class, by
considering the same dataset subset used to define \emph{V}. Bigrams that do
not belong to \emph{$V_2$} are mapped to the unknown bigram `$UNK_2$'. For each input file $F$ one can now build the feature vector $v_F$, which will
be the representation of $F$ in our model. To build this vector we first decode
$F$ from bytes to characters with the ASCII encoding and then tokenize the
result using vocabulary \emph{V}.  We then compute, for each token in \emph{V},
the frequency of that token among $F$'s tokens. We do the same for bigrams: for
each bigram in \emph{$V_2$}, its frequency among $F$'s tokens is determined.
Finally, we enumerate the elements of the set
$D = V \cup V_2 \cup \{`UNK'\} \cup \{`UNK_2'\}$ determined a fixed order for
them. The feature vector $v_F$ is built by assigning the computed frequency for
of $i$-th element of the ordered version of $D$ to the $i$-th component of the
vector itself. $v_F$ will represent the file $F$ in the following. Its
cardinality is $|D| = 5063$. This process is applied to each file in the
dataset; the resulting vectors will be used as inputs for the classification
algorithm. The same has been performed taking into account trigrams too but the best results have been achieved without them.

\subsection{Classifier}

As classifier we use a Deep Fully Connected Layers Neural Network which has
5063 input units, 133 output units
(which correspond to the possible extensions that we consider) and 3 hidden
layers with 1000, 800, and 700 units, respectively, with a dropout rate of 0.5
for each layer. The model structure is shown in Figure~\ref{fig:model}.

The classification problem is multi-class, single label. Hence we use in the
output layer the \emph{softmax} activation function 
$ \sigma\left(\vec x\right)_j=\frac{e^{x_j}}{\sum _{k=0}^{132} e^{x_k}}$ 
 which normalizes the values obtained from the previous layers with respect to the
available classes.
Output values represent the probabilities that a given file belongs to each
class. As we have multiple classes, we use the \emph{categorical cross entropy}
loss function. For each instance passed to the model, i.e. the $i$-th one,
 the loss function takes the form  $L\left(y_i,\hat y_i\right)= -\sum_{k=0}^{132}y_{i;k}ln\left(\hat y_{i;k}\right)$
where $y_i$ represents the actual ground truth label
(in the form of a one-hot vector) and $\hat y_i$ represents the
predicted probability output vector.
In the training phase we use the \emph{Adam} optimizer~\cite{kingma2014adam}
with learning rate $lr=0.0001$, which converge to good results (see
Section~\ref{sec:results}) after 8 epochs of training. During training the
model parameters were progressively modified in order to improve the similarity
of the predicted $\hat y_i$ vectors to the correspondent $y_i$ ground truth
vectors.

Given the relatively high number of classes in our problem, erroneous
classification is likely to happen. To investigate it we keep track of
classification errors (on the validation set) in two different ways. We will
use the following assumptions and notations: $C$ is the set of classes, $Vset$
denotes our validation set, while for $x\in Vset$, $GroundT(x) = i$ iff $i$ is
the ground truth label of $x$ and $Predict(x) = i$ iff $i$ is the label
assigned by the classifier to $x$. Moreover, when we perform a prediction for
an input $x\in Vset$ we obtain a probability value $p_x(i)$ for each possible
class $i\in C$.  The predicted class is the one which has the highest
probability value, that is, $Predict(x) = m$ iff $p_x(m) \geq p_x(j)$ for each
$j\neq m$, $j\in C $. First, we measure how many of the examples of a given class are classified in a
wrong way. This is done by introducing, for each pair of classes $i$ and $j$, a
quantity $T_{ij}$ indicating how many times a file whose label (i.e., ground
truth) is $i$ is classified in the class $j$. More precisely we define
$T_{ij} =\frac{ \sum_{x\in Vset } 1 \mid GroundT(x) = i \hbox{ and }
    Predict(x) = j}{\sum_{x\in Vset } 1 \mid GroundT(x) = i } $.

A second way to register classification errors consists in considering also the
second, third, fourth, and fifth best choices for classifying an example and
see whether some of them significantly co-occur with the predicted class. More
precisely, given $x\in Vset$, assume that $Predict(x) = m$ and that
$pr_x(m) \geq p_x(h) \geq p_x(k) \geq p_x(l) \geq p_x(r) \geq p_x(i)$, for
$h, k, l, r\in C$ and for each other $i \in C$. In this case we say that
$m, h, k, l, r$ are the top five classifications for $x$, written
$Top(x) =\{m, h, k, l, r\}$, for short. Then we normalise these five values by
defining
  $pn_x(m)=\frac{pr_x(m)}{p_x(m) + p_x(h) + p_x(k)+ p_x(l) + p_x(r)}$
and analogously for the other four values. 

Then, for each pair of classes $i$
and $j$, we define
$S'_{ij} =\sum_{x\in Vset} pn_x(j) \mid Predict(x) = i \hbox{ and } j \in
  Top(x) $ and we normalize this value by considering by the total number of
times in which the i-th class has been predicted, as follows:
$S_{ij} = \frac{S'_{ij}} {\sum_{x\in Vset} 1 \mid Predict(x) = i}$.

Given these quantities $T_{ij}$ and $S_{ij}$ we set two thresholds for their
values, $\tau_T$ and $\tau_S$ respectively. Given a pair of classes $i$ and
$j$, if $T_{ij}>\tau_T$ or $S_{ij}>\tau_S$ we consider the classes $i$ and $j$
somehow related in the predictions and we say that they belong to the same
``confusion group''. By setting the thresholds $\tau_T= 0.05$ and
$\tau_S= 0.02$ we obtain the confusion groups shown in Table~\ref{tab:conf}.
The labels which do not appear in the table are those that do not pass the
thresholds, meaning that the classifier do not incur into relevant ambiguity
for them.

\begin{table}
  \caption{Relevant extension confusion groups. Note how some groups contain
    labels that commonly refer to the same or similar file types, justifying
    the origin of the ambiguity.}
  \label{tab:conf}
  \begin{tabular}{l|l|l|l}
    \toprule
    Group ID & Extensions&Group ID & Extensions\\
    \midrule
    $0$ & \verb .bash , \verb .sh , \verb .ps1 , \verb .after , \verb .jet , \verb .kt , \verb .template &
    $1$ & \verb .markdown , \verb .md \\
    $2$ & \verb .cmake , \verb .cmd , \verb .yaml  , \verb .yml  , \verb .rst , \verb .txt , \verb .baseline , \verb .bat &
    $3$ & \verb .dts , \verb .dtsi \\
    $4$ & \verb .ctp , \verb .php &
    $5$ & \verb .ml , \verb .mli \\
    $6$ & \verb .csproj , \verb .ilproj &
    $7$ & \verb .jl , \verb .j \\
    $8$ & \verb .rb , \verb .cr , \verb .exs &
    $9$ & \verb .h , \verb .ino , \verb .hpp \\
    $10$ & \verb .m4 , \verb .ac &
    $11$ & \verb .cpp , \verb .cc \\
    $12$ & \verb .clj , \verb .cljs &
    $13$ & \verb .swift , \verb .sil , \verb .gyb \\
    $14$ & \verb .tsx , \verb .jsx , \verb .js , \verb .ts , \verb .htm , \verb .html &
    $15$ & \verb .cjsx , \verb .coffee \\
    $16$ & \verb .css , \verb .scss &
    $17$ & \verb .after  , \verb .kt \\
    \bottomrule
  \end{tabular}
\end{table}

\section{Results}
\label{sec:results}

The architecture underlying the proposed model is quite simple in comparison to
other machine learning approaches used to treat text. Contrarily to complete
automatic feature learning algorithms, such as those used in Convolutional,
Recursive, or Attention Neural Networks, computations in our model are faster,
both for training and prediction. In particular, the learning task can be
completed in a relatively short time: it took around 10 hours of training for 8
epochs to train the
parameters of the model and to reach $\approx 85\%$ of accuracy on the validation and test sets. Predictions are made
transforming the input feature vectors by means of simple operations such as
vector multiplications, sums, and activation functions applications based on
the parameters learned during the training phase.
The test set consists of $\approx 2M$ elements that we extracted from the
original dataset after the pre-processing described in
Section~\ref{sec:dataset} except for the balancing step as this could bring to
unfair evaluations and the resulting set would not represent the real world statistical
distribution of classes, excluding of course the elements that were used in the
train and validation sets.
Various performance measures are used here and they are all based on the
confusion matrix which is computed on the test set and whose generic element
$C_{ij}$ contains the number of files of class (extension) $i$ which are
classified as $j$. The average accuracy value obtained on all the classes by
the model is $85\%$.
We report in Table~\ref{tab:scores} the values for
precision
 $P_i=\frac{C_{ii}}{\sum_{j=0}^{120}Cji}$, recall  $R_i=\frac{C_{ii}}{\sum_{j=0}^{120}Cij}$ and $F_1$-score
$F_i=\frac{2R_iP_i}{R_i+P_i}$ for each of the considered classes. The extensions are shown in descending order of frequency within the original test set.

\begin{longtable}{l|ccc"l|ccc"l|ccc}
  \caption{Performance measures for the encoder architecture on the testset (2019)%
    \label{tab:scores}
  }\\
  \hline
  \textbf{Ext} & \textbf{P} & \textbf{R} & \textbf{F} &\textbf{Ext} & \textbf{P} & \textbf{R} & \textbf{F} &\textbf{Ext} & \textbf{P} & \textbf{R} & \textbf{F} \\
  \hline
  \endhead
  \texttt{.js} & 0.93 & 0.69 & 0.79 &\texttt{.vb} & 0.95 & 0.99 & 0.97& \texttt{.gyp} & 0.32 & 1.0 & 0.48\\
  \texttt{.c} & 0.96 & 0.94 & 0.95&\texttt{.asciidoc} & 0.64 & 0.95 & 0.76 & \texttt{.bat} & 0.42 & 0.72 & 0.53\\
  \texttt{.html} & 0.98 & 0.87 & 0.92&\texttt{.gradle} & 0.75 & 0.95 & 0.84 &\texttt{.erl} & 0.6 & 1.0 & 0.75 \\
  \texttt{.java} & 0.99 & 0.97 & 0.98&\texttt{.cr} & 0.34 & 0.92 & 0.5 & \texttt{.gemspec} & 0.78 & 1.0 & 0.88 \\
  \texttt{.h} & 0.93 & 0.71 & 0.81&\texttt{.lua} & 0.7 & 0.95 & 0.81 & \texttt{.fish} & 0.58 & 0.95 & 0.72  \\
  \texttt{.py} & 0.99 & 0.93 & 0.96 &\texttt{.ex} & 0.83 & 0.96 & 0.89 &\texttt{.i} & 0.09 & 0.89 & 0.16\\
  \texttt{.go} & 0.99 & 0.96 & 0.97 & \texttt{.ilproj} & 0.79 & 0.95 & 0.86& \texttt{.texi} & 0.93 & 1.0 & 0.96 \\
  \texttt{.md} & 0.97 & 0.72 & 0.83 & \texttt{.dtsi} & 0.64 & 0.88 & 0.74 &\texttt{.template} & 0.09 & 0.48 & 0.15\\
  \texttt{.rb} & 0.98 & 0.85 & 0.91&\texttt{.props} & 0.79 & 0.97 & 0.87 &\texttt{.pl} & 0.62 & 0.95 & 0.75 \\
  \texttt{.json} & 0.95 & 0.95 & 0.95& \texttt{.vcxproj} & 0.97 & 0.99 & 0.98 &\texttt{.ac} & 0.68 & 0.99 & 0.81 \\
  \texttt{.cpp} & 0.74 & 0.59 & 0.66& \texttt{.clj} & 0.96 & 0.99 & 0.97 & \texttt{.groovy} & 0.23 & 0.91 & 0.37 \\
  \texttt{.ts} & 0.65 & 0.82 & 0.73&\texttt{.markdown} & 0.13 & 0.73 & 0.22 &\texttt{.mak} & 0.86 & 0.96 & 0.91  \\
  \texttt{.php} & 0.96 & 0.89 & 0.92&\texttt{.symbols} & 0.88 & 0.99 & 0.93 &\texttt{.vbproj} & 0.54 & 0.98 & 0.7\\
  \texttt{.cs} & 0.98 & 0.96 & 0.97 & \texttt{.hs} & 0.78 & 0.99 & 0.87 &\texttt{.pkgproj} & 0.85 & 0.98 & 0.91\\
  \texttt{.rs} & 0.97 & 0.97 & 0.97&\texttt{.dts} & 0.75 & 0.69 & 0.72& \texttt{.sql} & 0.16 & 0.93 & 0.27 \\
  \texttt{.cc} & 0.58 & 0.77 & 0.66&\texttt{.el} & 0.95 & 0.99 & 0.97 &\texttt{.j} & 0.22 & 0.98 & 0.36\\
  \texttt{.xml} & 0.96 & 0.93 & 0.94& \texttt{.proto} & 0.53 & 0.99 & 0.69&\texttt{.tpl} & 0.12 & 0.77 & 0.21  \\
  \texttt{.glif} & 1.0 & 1.0 & 1.0& \texttt{.toml} & 0.6 & 0.98 & 0.74& \texttt{.rake} & 0.1 & 0.95 & 0.18  \\
  \texttt{.txt} & 0.82 & 0.52 & 0.64 &\texttt{.pbxproj} & 0.98 & 1.0 & 0.99 & \texttt{.textile} & 0.2 & 0.99 & 0.33 \\
  \texttt{.kt} & 0.96 & 0.72 & 0.82& \texttt{.exs} & 0.56 & 0.95 & 0.7 &\texttt{.webidl} & 0.59 & 0.99 & 0.74  \\
  \texttt{.scala} & 0.96 & 0.95 & 0.95 & \texttt{.mk} & 0.79 & 0.91 & 0.85&\texttt{.bash} & 0.22 & 0.77 & 0.34 \\
  \texttt{.swift} & 0.92 & 0.85 & 0.88& \texttt{.sil} & 0.71 & 0.97 & 0.82& \texttt{.cjsx} & 0.35 & 0.97 & 0.51 \\
  \texttt{.yml} & 0.77 & 0.81 & 0.79 & \texttt{.after} & 0.19 & 0.74 & 0.3& \texttt{.pb} & 0.39 & 1.0 & 0.56  \\
  \texttt{.css} & 0.88 & 0.9 & 0.89 & \texttt{.erb} & 0.19 & 0.77 & 0.3&\texttt{.builds} & 0.93 & 1.0 & 0.96\\
  \texttt{.rst} & 0.59 & 0.89 & 0.71 & \texttt{.jade} & 0.27 & 0.9 & 0.42&\texttt{.vcproj} & 0.92 & 1.0 & 0.96 \\
  \texttt{.sh} & 0.84 & 0.83 & 0.83&  \texttt{.gyb} & 0.33 & 0.97 & 0.49&\texttt{.xcscheme} & 1.0 & 1.0 & 1.0  \\
  \texttt{.csproj} & 0.96 & 0.9 & 0.93 &\texttt{.log} & 0.57 & 0.91 & 0.7&\texttt{.ngdoc} & 0.1 & 0.96 & 0.18 \\
  \texttt{.coffee} & 0.82 & 0.92 & 0.87&\texttt{.ipynb} & 0.43 & 0.98 & 0.6 &\texttt{.perl} & 0.66 & 0.98 & 0.79 \\
  \texttt{.scss} & 0.85 & 0.93 & 0.89& \texttt{.cmake} & 0.29 & 0.95 & 0.44&\texttt{.eslintrc} & 0.13 & 0.98 & 0.23 \\
  \texttt{.phpt} & 0.68 & 0.94 & 0.79&\texttt{.ps1} & 0.78 & 0.96 & 0.86&\texttt{.sbt} & 0.61 & 0.97 & 0.75 \\
  \texttt{.jsx} & 0.26 & 0.79 & 0.39 &\texttt{.pyx} & 0.43 & 0.95 & 0.59 &\texttt{.handlebars} & 0.1 & 0.97 & 0.18\\
  \texttt{.hpp} & 0.2 & 0.83 & 0.32 &\texttt{.tmpl} & 0.24 & 0.67 & 0.35&\texttt{.iml} & 0.62 & 1.0 & 0.77\\
  \texttt{.xht} & 0.79 & 0.97 & 0.87 &\texttt{.m4} & 0.74 & 0.95 & 0.83&\texttt{.rml} & 0.79 & 1.0 & 0.88\\
  \texttt{.tsx} & 0.34 & 0.81 & 0.48&\texttt{.check} & 0.24 & 0.81 & 0.37 &\texttt{.cmd} & 0.28 & 0.81 & 0.42\\
  \texttt{.jl} & 0.9 & 0.79 & 0.84&\texttt{.il} & 0.77 & 1.0 & 0.87 &\texttt{.zsh} & 0.2 & 0.94 & 0.33 \\
  \texttt{.htm} & 0.41 & 0.93 & 0.57&\texttt{.am} & 0.54 & 0.96 & 0.69&\texttt{.tcl} & 0.65 & 0.98 & 0.78  \\
  \texttt{.dart} & 0.74 & 0.98 & 0.84 &\texttt{.adoc} & 0.58 & 0.95 & 0.72&\texttt{.xib} & 0.55 & 1.0 & 0.71   \\
  \texttt{.ml} & 0.98 & 0.95 & 0.96&\texttt{.mli} & 0.68 & 1.0 & 0.81&\texttt{.jet} & 0.05 & 0.91 & 0.09    \\
  
  \texttt{.m} & 0.72 & 0.95 & 0.82& \texttt{.sln} & 0.99 & 1.0 & 0.99 &\texttt{.dsp} & 0.94 & 1.0 & 0.97  \\
  
  \texttt{.haml} & 0.92 & 0.98 & 0.95& \texttt{.sass} & 0.69 & 0.96 & 0.8 &\textbf{\small micro avg.} & 0.85 & 0.85 & 0.85 \\
  \texttt{.yaml} & 0.46 & 0.65 & 0.54&\texttt{.w32} & 0.31 & 0.99 & 0.47 &\textbf{\small macro avg.} & 0.64 & 0.91 & 0.71 \\

  \bottomrule
\end{longtable}


We obtained good overall results and we report at the end of the table also micro- and macro-average. The former is
useful in order to take into account the number of instances per class: classes
with the higher number of examples will have a heavier influence in the average
value than the less popular ones the. In fact, micro-average is
defined as  $P_m=\frac{\sum_{i}TP_i}{\sum_i\left(TP_i+FP_i\right)}$
where $TP_i$ are the true positives and $FP_i$ are the false positives .
Macro-average is defined without taking into account the number of instances
per class, i.e., every class will equally influence the average.
It appears that the main problem faced by the model is the presence of very
similar classes that could actually be treated as the same class, for example
both \texttt{.yaml} and \texttt{.yml} are used for files written in YAML, which
can introduce errors in the predictions.  Also, there are classes whose text
contents can be very general and therefore could not be easily recognized, such
as files with the \texttt{.txt} extension.

\section{Threats to Validity}
\label{sec:threats}

{\em External validity.}
Even if the proposed classifier has been designed and tested on a fairly large
set of files, the used dataset falls short of the entire corpus of source code distributed via
publicly accessible version control systems. Datasets that approximate that
corpus~\cite{swh-cacm-2018, swh-ipres-2017, mockus2019woc} do exist and could
potentially be more challenging to tackle because: (1) they will include more
extension/classes that did not occur within this paper datasets, (2) they can
include noisier data as repository popularity is likely a proxy of project
quality in terms of coding practices, and (3) they can include files from more
varied chronological epochs---also programming languages always evolve and new
ones emerge increasing the variety of classes to be considered. 
Nevertheless our design choices were oriented to address this kind of problems, 
as we discussed before, and we aim to
use the proposed approach on these larger datasets as future work.

{\em Internal validity.}
Various (non domain-specific) heuristics have been applied. First, filtering of
non-textual files has been performed based on the percentage of non-printable
characters within a random sample of the datasets; different samples or
thresholds might affect stated results. On the same front, the choice of using
the ASCII encoding to detect printable characters due to the lack of encoding
information. An alternative approach would have been guessing the used encoding
(using libraries such as \texttt{chardet}); it is not clear which biases either
approach would introduce, if any.

The vocabularies \emph{V} and \emph{$V_2$} are based on tokens/bigrams
frequency distributions defined separately on different classes. This can
introduce model biases and could be mitigated by using separate datasets to
define the vocabularies and to train the neural network. Hyper parameters tuning has been performed as an iterative process, certainly
not exhaustive. In spite of the achieved good results it is possible that
different choices for hyper parameters and/or neural network topology would
score even better. At  the end of previous section we mentioned some possible threats due 
to extensions classification, in particular
some of the extension classes that we treated as different might actually refer
to the same abstract file type, which can cause confusion for the model and it
makes the training task harder as a class share its instances with other
classes. We maintain it is difficult to mitigate this last threat without
relying on domain-specific knowledge that we wanted to avoid using, at least as
a first approximation in the present study.

\section{Related Work}
\label{sec:related}
\paragraph{Big code}

Information retrieval~\cite{frakes1992ir} have been applied extensively to all
sorts of software development artifacts, including source code. The most active
and recent trend in the field goes under the label of ``big code'' and consists
in the application of machine learning techniques to software development
artifacts. In the remainder of this section we focus on related work in the
area of file type and programming language detection. For big code work outsite
that field of application we refer to the work of Allamanis et
al.~\cite{allamanis2018survey} that surveys an impressive number of
applications and approaches in the field.

{\em Language detection.} Several approaches have explored for for programming language detection.  van
Dam and Zaytsev~\cite{van2016software} tested various programming language
classifiers on source files extracted from GitHub for 19 language classes and
libelling them using Linguist as a source of truth. They obtained a value of
0.97 for $F_1$-score, precision, and recall. Klein et al.~\cite{klein2011algorithmic} performed language recognition among
25 language classes using source code from GitHub and labeling files using both
file extensions and the GitHub language detection tool (Linguist discussed
below). Only files for which file extension and Linguist-detected language
match have been included in their trainset. Then, a feature vector is extracted
from source code using features such has parentheses used, comments, keywords,
etc., and used for training. The obtained accuracy is 50\%. Ugurel et al.~\cite{ugurel2002classification} proposed various mechanisms for
the classification of source code, programming language, and topics, based on
support vector machines (SVM). On the language front they were able to
discriminate among 10 different programming languages with 89\% accuracy.
Their data were retrieved from various source code archives available at the
time (it was 2002, pre-GitHub). With respect to the aforementioned studies the approach presented in this paper
is simpler, performs better in terms of accuracy, and handles significantly
more (5--10x) file type classes. Reyes et al.~\cite{reyes2016automatic} used a long short-term memory (LSTM)
algorithm to train and test a programming language classifier on 3 different
datasets, one from Rosetta Code~\cite{nanz2015rosetta} (391 classes), GitHub
archive (338), and a custom dataset (10). The obtained accuracies were,
respectively, 80\%, 29\%, and 100\%. They also compared their results to
Linguist, finding Linguist scored worse except for the second dataset in which
it reached 66\% accuracy.
The comparison between Rayes et al.~and the approach presented in the present
paper is interesting. The custom dataset confirms what was already apparent
from previous comparisons: one can do much better in terms of accuracy reducing
the number of language classes (and as little as 10 classes is not enough for
our stated purpose). The other two datasets exhibit larger diversity than ours
($\approx$\,300 classes v.$\approx$\,100), but perform very differently. We
score better than the (more controlled) Rosetta Code dataset and \emph{much}
better than the GitHub dataset. Our approach is simpler in terms of
architecture than theirs and we expect it to perform better in terms of
training and recognition time (as LSTM tends to be slow to train)---but we have
not benchmarked the two approaches in comparable settings so this remains a
qualitative assessment at this stage. Linguist~\cite{linguist} is an open source language detection tool developed by
GitHub. Its own accuracy is reported by GitHub~\cite{linguist-accuracy} as
being around 85\%. The accuracy of studies that have used Linguist as ground
truth should then be diminished accordingly. Additionally, Linguist relies on
file extensions as a feature and its accuracy drops significantly when they
are missing~\cite{linguist-accuracy}, as it is the case in our problem
statement.
The work by Gilda~\cite{gilda2017source} is very similar to ours. File
extensions are used as ground truth for source code extracted from GitHub and a
word-level convolutional neural network (CNN) is used as classifier. The model
reached 97\% accuracy and is able to classify 60 different languages. We
perform significantly better in terms of diversity (2x classes), but slightly
worse in terms of accuracy (-8\%).

{\em File-type detection}. Previous work has also been done on the more general problem of classifying
file types that might also be binary formats. In the field of digital forensic
Fitzgerald et al.~\cite{fitzgerald2012using} performed classification for 24
file classes using byte-level 1-grams and bigrams to build feature vectors fed
to a SVM algorithm, reaching 47\% accuracy on average.
Gopal et al.~\cite{gopal2011statistical} used and compared similar approaches
for the same task, but for 316 classes including 213 binary formats and 103
textual ones. They reached 90\% micro-average and 60\% macro-average
$F_1$-score. This relatively big gap between the two average measures hints at
significant differences in the classes (e.g., frequencies in the test set).
Binary file type detection is a significantly different problem than
ours. There one can rely on file signatures (also known as ``magic numbers''),
as popular Unix libraries like \texttt{libmagic} and the accompanying
\texttt{file} utility do. Such approaches are viable and could be very
effective for binary files, but they are less maintainable in the long run (as
the database of heuristics should be maintained lest it becomes stale) and less
effective on textual file formats, where magic numbers are either missing or
easily altered.

\section{Conclusion}
\label{sec:conclusion}

In this work we considered the problem of predicting textual file types for
files commonly found in version control systems (VCS), relying solely on file
content without any a priori domain knowledge or predetermined heuristic. The
problem is relatively novel (as most existing language/file type detectors rely
heavily on extensions) and relevant in contexts where extensions are missing or
cannot be trusted, and shed light into the intrinsic recognizability of textual
files used in software development repositories. We propose a simple model to solve the problem based on an universal word
tokenizer, word-level n-grams of length 1 to 2, feature vectors based on n-gram
frequency, and a fully connected neural network as classifier. We applied the
model to a large dataset extracted from GitHub spanning 133 well represented
file type classes. In spite of its simplicity the model performed very well,
nearing 85\% average accuracy, and outperforming previous work in either
accuracy, number of supported classes, or both. We expect that model simplicity
will make it more maintainable in the future and less computationally expensive
to train and run than alternatives based on learning algorithms such as CNN,
LSTM, or Attention.

Concerning future work, a straightforward next step is scaling up experimentation of the proposed
model, moving from the high-starred GitHub dataset we used for this work to
larger and more diverse datasets such as Software
Heritage~\cite{swh-cacm-2018}. In that context we will have significantly more
starting classes and hopefully enough samples in each of them for enlarging the
set of labels actually used in training.
As we observed, extensions alone are ambiguous in many cases and this poses
challenges in training and evaluation. To mitigate this issue it is worth
exploring the possibility of inserting narrow domain knowledge about file
extensions that often go together. It is not clear whether doing so, partly
backtracking the ``no domain knowledge'' assumption of this work, would be
worth the effort in terms of increased accuracy; hence, it is worth exploring.
An alternative approach for improving over the current handling of model
confusion is adding a second tier of classifiers, one for each class of
ambiguous extensions, a popular technique in NLP. There are various methods to
combine the predictions from the first and second level classifiers which
should be explored. We have briefly explored the topic of accuracy degradation in the lack of
retraining at a 2-year distance. A more general characterization would be
interesting to have and is feasible to obtain it by exploiting historical
software archives and/or VCS timestamp information. Such a characterization
will allow to devise data-driven approaches for when and how to retrain file
type and language detection tools in a world in which programming languages
constantly evolve.

\begin{acks}
  The authors would like to thank Éric de La Clergerie and Benoit Sagot for
  insightful discussions on automated techniques for natural language
  detection, without which this work would not have been possible.
\end{acks}
 
\bibliographystyle{ACM-Reference-Format}
\bibliography{main}
  
\end{document}